\documentclass[superscriptaddress,twocolumn,showpacs]{revtex4}
\usepackage{mathrsfs}
\usepackage{amssymb}
\usepackage[tbtags]{amsmath}
\usepackage{graphicx}
\usepackage{epsfig,graphicx,times}
\usepackage{color}
\usepackage{subfigure}

\begin{document}
\title{Deterministic generations of NOON states via shortcuts to adiabaticity}
\author{Jingwei Chen}
\affiliation{State Key Laboratory of Optoelectronic Materials
and Technologies, School of Physics, Sun Yat-sen
University, Guangzhou 510275, China}

\author{L. F. Wei\footnote{E-mail: weilianfu@gmail.com}}

\affiliation{State Key Laboratory of Optoelectronic Materials
and Technologies, School of Physics, Sun Yat-sen
University, Guangzhou 510275, China}
\affiliation{Quantum Optoelectronics Laboratory, School of Physics and Technology, Southwest Jiaotong University, Chengdu 610031, China}

\begin{abstract}
NOON states play the important roles in quantum information processings and quantum metrology, but the fidelities of these states generated previously are limited typically by the practically-unavoidable decoherence and operational imperfections.
Here, we propose an efficient scheme to generate photonic NOON states alternatively by rapid population passage technique via shortcut to adiabaticity (STA), rather than the usual Rabi oscillations. Since the deterministic population passages based on the STAs are insensitive to details of the operations and can be implemented as fast as the Rabi oscillations, the fidelity of the generated NOON state could be satisfactorily high. The feasibility of the proposal is demonstrated specifically with the experimental circuit QED systems by rapidly driving two artifical qutrits.

\end{abstract}


\maketitle

{\it Introduction.---}It is well-known that quantum entanglement lies at the core of quantum physics. In particular, the so-called NOON states $|\psi\rangle=(|N0\rangle+e^{i\phi}|0N\rangle)/\sqrt{2}$ (with arbitrary phase $\phi$ and $N=0,1,2,...$) have drawn more and more attention~\cite{N1,N2,N3,N4,N5,N6,N7,N8,N9,N10,N11,N12,N13}, due to their potential widely applications in quantum precise measurements~\cite{N1,N2,N3}, quantum lithography~\cite{N4}, quantum information processings~\cite{N14}, and quantum communication~\cite{N15}, etc..
Especially, with the NOON states~\cite{N1,N2,N3} the precisions of certain measured quantities can be greatly improved from the standard quantum limit $1/\sqrt{N}$ to the Heisenberg limit $1/N$. However, the experimental generations the desired NOON states with satisfactorily high fidelities for these applications is still a challenge.

So far, two approaches have usually being utilized to create the NOON states; post-selected measurements based beam-splittings for traveling photons and rapid Rabi oscillations in solid-state systems. Theoretically~\cite{N5}, the ideal photonic NOON states with arbitrary $N$ can be generated by employing the interference of a classical coherent light and a quantum light produced by spontaneous parametric down-conversion (SPDC) through a $50:50$ beam splitter. However, in practice, the fidelities of the generated photonic NOON are limited by the finite setup transmission coefficient $\eta$ and thus are always less than unit, e.g., $72.1\%$ NOON state with $N=4$ for $\eta=0.11$~\cite{N6}.
Alternatively, the NOON states can be generated by quantum state population engineering~\cite{N7,N8,N9,N10,N11,N12,N13}; using the rapid Rabi oscillation to implement sequentially the population transfers from the initial ground state $00$ to the expected NOON state with $N>0$. However, due to the inevitable decoherence and the imperfect operations, the fidelities of the generated NOON by this approach are still very limited; typically, e.g., $33\%$ of the $N=3$ NOON state for the standing wave photons in circuit quantum electrodynamics (QED) system~\cite{CQED1,CQED2}.
Therefore, it is greatly hope to find a new approach to generate the desired high-fidelity NOON states.

Given the finite coherence time of the current manipulatable quantum system is always limited, a potential scheme to generate the high-fidelity NOON state could be achieved by optimizing the population transfers, e.g., realizing the rapid population passages deterministically via decreasing various operational imperfections. Although the Rabi oscillation (RO) technique has being used widely to rapidly control the population transfers between the selected quantum states, the fidelities of these transfers are quite sensitive to any parameter fluctuations; the integral area and the frequency variation of the applied pulses.
While, the so-called adiabatic population passage (APP) technique~\cite{AP1,AP2,AP3,AP4} is insensitive to the details of the operational pulses, but the adiabatic condition required in this technique limits practically the fidelities of the implemented population transfers.
In order to overcome such a difficulty, in this letter we propose a modified scheme to generate high-fidelity photonic NOON states deterministically by using the shortcut to the adiabaticity (STA) technique~\cite{STA1,STA2,STA3,STA4,STA5,STA6,STA7}. We find that, by applying certain suitable additional counter-diabatic fields, the desired NOON states for two standing wave fields in two cavities could be generated rapidly beyond the adiabatic limit. It is shown that the high-fidelity for each population passages for the desired generation is significantly higher than that by RO technique. Specifically, our proposal is demonstrated with the experimentally-existing devices, i.e., the circuit QED system, by driving two interacting three-level atoms.

{\it Robust generation of photonic NOON states via STA technique.---}
We consider a quite generic model, i.e., two coupled ladder-type three-level atoms (called hereafter as the qutrits 1 and 2, respectively) interact respectively with two individual cavities $a$ and $b$.
As a beginning, we assume that two cavities are cooled to their vacuums and the two qutrits stay at their ground states, i.e., the initial state of the system is $|g_1g_20_a0_b\rangle$. Then the desired NOON state of the cavities can be implemented generically by the following population transfers:

i)Keep the states the cavities unchanged but drive the qutrits into the Bell state, i.e., the state of the system becomes as
\begin{equation}
|\psi_0\rangle=\frac{1}{\sqrt{2}}(|e\rangle_1|g\rangle_2+|g\rangle_1|e\rangle_2)\otimes|0_a0_b\rangle,
\end{equation}
ii)Transfer the population of the excited state $|e\rangle$ of each qutrits into
its auxiliary state $|f\rangle$, keeping the states of the cavities still unchanged.
This let the state of the two qutrits become
$\frac{1}{\sqrt{2}}(|f\rangle_1|g\rangle_2+|g\rangle_1|f\rangle_2)$.
iii) Swap the population of the auxiliary state of each qutrits to the coupled cavity, and evolve the system to the state
\begin{equation}
|\psi_1\rangle=\frac{1}{\sqrt{2}}(|e\rangle_1|g\rangle_2|1\rangle_a|0\rangle_b+|g\rangle_1|e\rangle_2|0\rangle_a|1\rangle_b).
\end{equation}
iv) Repeat the above steps $N-1$ times, and then return the qutrits to their ground states. This evolves finally the system to the state~\cite{N8}
\begin{equation}
|\psi_N\rangle=\frac{1}{\sqrt{2}}|g\rangle_1|g\rangle_2(|N\rangle_a|0\rangle_b+|0\rangle_a|N\rangle_b),
\end{equation}
with the desired photonic NOON state of the two cavities being generated.

The generation demonstrated above seems quite trivial and can be simply implemented by the usual RO technique. Beside the usual decoherence of the system which needs to increase the quantum quality of the system essentially, the relevant fidelity is limited mainly by the following operational imperfections: a) the frequency deviations of the target resonant-RO pluses and b) the imprecisions of the integral area of the applied pulses due to the fluctuations of the strengths and durations of the pulses. This is the reason that the fidelity of the NOON state generated previously by the RO technique is quite finite. In principle, these difficulties could be effectively overcome by the replacing the RO pluses by the APP ones~\cite{AP1,AP2,AP3}, which keeps always the system in the instantaneous eigenstates $|\lambda_n(t)\rangle$ of the time-dependent Hamiltonian $H(t)$ of the driven system. By properly setting the parameters of the pulses the designed population passages can be implemented deterministically~\cite{AP1,AP2,AP3}, as long as the relevant adiabatic conditions are satisfied robustly.
Physically, the adiabatic condition implies that the population passage desired above should be sufficiently slow (as any nonadiabatic transition influences the passage fidelity), but the duration is still required to be shorter significantly than the finite decoherence time of the system.
Fortunately, with the help of the STA~\cite{STA1,STA2,STA3,STA4,STA5}, such a contradiction can be solved. The key point of STA is to add an auxiliary counter-diabatic field to avoid any potential non-adiabatic transition during the above adiabatic process. In fact, the required counter-diabatic field can be designed by various methods, e.g., the transitionless driving algorithm~\cite{STA1,STA2,STA3}, counterdiabatic control protocols~\cite{STA4}, and Lewis-Riesenfeld invariants theory~\cite{STA5}, etc.. For example, with the transitionless driving algorithm~\cite{STA1}, one can apply a counter-diabatic driving
$H_1(t)=i\hbar\sum_n(|\partial_t\lambda_n(t)\rangle\langle\lambda_n(t)|
-\langle\lambda_n(t)|\partial_t\lambda_n(t)\rangle|\lambda_n(t)\rangle\langle\lambda_n(t)|)$
to robustly keep the system $H(t)$ at its instantaneous eigenstate $|\lambda_n(t)\rangle$ without any unwanted transition to all the other instantaneous eigenstates $\{|\lambda_m(t)\rangle, m\neq n\}$. Based on this idea, we show below how to generate NOON states with two cavities by coupling them to two manipulatable qutrits, i.e., implementing the operations $i)$ to $iii)$ demonstrated above. The Hamiltonian of the system originally reads ($\hbar=1$)
\begin{equation}
H=H_q+H_r+H_i,
\end{equation}
with $H_r=\omega_aa^{\dag}a+\omega_bb^{\dag}b$ and $H_q=\sum_{i=1,2}\omega_i^g|g\rangle_i\langle g|_i+\omega_i^e|e\rangle_i\langle e|_i+\omega_i^f|f\rangle_i\langle f|_i)+g_{12}|g,e\rangle\langle e,g|+g_{12}'|e,f\rangle\langle f,e|+H.c.$
being respectively the Hamiltonians of the cavities and qutrits, and
\begin{eqnarray}
H_i=\sum_{i=1,2}(g_i|g_i\rangle\langle e_i|+g_i'|e_i\rangle\langle f_i|)a^{\dag}_i
+H.c.,
\end{eqnarray}
the qutrit-cavity interactions with $a_1=a,\,a_2=b$.
For simplicity, the qutrit-qutrit interacting strength $g_{12}$ and the qutrit-cavity coupling one $g_i,\,g'_i$ are assumed to be tunable.

First, to perform the i) operation, i.e., prepare the two qutrits in the Bell states, by the STA technique, we switch off the qutrit-qutrit and qutrit-cacity couplings and then deterministically excite one of the qutrits. For example, a time-dependent driving
\begin{equation}\label{STA1}
H_1(t)=\frac{\hbar}{2}[\Omega_1(t)(|e\rangle_1\langle g|_1+|g\rangle_1\langle e|_1)+2\Delta_1(t)|e\rangle_1\langle e|_1],
\end{equation}
is applied to the qutrit 1. If the two pulses $\Omega_1(t)$ and $\Delta_1(t)$ are chosen properly to tune the mix angle $\theta_1(t)=\arctan[\Omega_1(t)/\Delta_1(t)]/2$ from zero to $\pi/2$, then the state of the system is driven from the state $|g,g,0,0\rangle$ to the $|e,g,0,0\rangle$. In order to eliminate all the possible diabatic transitions during such a process, an additional driving
\begin{equation}\label{STA1a}
H_1'=\frac{\hbar}{2}\Omega_1'(t)(i|e\rangle_1\langle g|_1-i|g\rangle_1\langle e|_1),
\end{equation}
with
$\Omega_1'(t)=[\Omega_1(t)\dot{\Delta}_1(t)-\dot{\Omega}_1(t)
\Delta_1(t)]/[\Omega_1^2(t)+\Delta_1^2(t)]$ is required to apply simultaneously.
Many pluses, typically the usual Allen-Eberly (AE) drivings~\cite{AE}:
$\Omega_1(t)=\Omega_0\textrm{sech}(\pi t/2t_0),\,\,
\Delta_1(t)=(2\beta^2t_0/\pi)\tanh(\pi t/2t_0)$,
can be used to rapidly implemented such a process beyond the adiabatic limit.
Now, we want to rapidly swap the population of the qutrit 1 to qutrit 2. Note that the qutri-qutrit interacting Hamiltonian reads  writes
\begin{eqnarray}
H_2(t)&=&\frac{\hbar}{2}[g_{12}(|e,g\rangle\langle g,e|+|g,e\rangle\langle e,g|)\nonumber\\
&+&2\Delta_{12}(t)|g,e\rangle\langle g,e|],\,\,\Delta_{12}(t)=\omega_1'-\omega_2'
\end{eqnarray}
in the invariant space $\{|e,g\rangle,|g,e\rangle\}$. Thus, one can
choose the proper pulse series, e.g.,
\begin{eqnarray}
g_{12}&=&G_0\exp(-(t-\tau)^2/T_0^2),\\
\Delta_{2}&=&\Delta_0\exp(-(t+\tau)^2/(mT_0)^2),
\end{eqnarray}
to tune the mix angle $\theta_2(t)=\arctan[g_{12}/\Delta_{12}(t)]/2$ from zero to $\pi/4$.
After this process, the state of system will adiabatically evolve from the state $|e,g,0,0\rangle$ to the Bell state $\frac{1}{\sqrt{2}}(|e,g,0,0\rangle+e^{(i\phi)}|g,e,0,0\rangle)$, with the removable phase factor $\phi$. Again, in order to avoid the possible adiabatic transition during this population trasfer, a counter-diabatic driving
\begin{equation}
H_2'(t)=\frac{i\hbar J(t)}{2}(|e,g\rangle\langle g,e|-|g,e\rangle\langle e,g|)
\end{equation}
with
$
J(t)=[g_{12}(t)\dot{\Delta}_{12}(t)-\dot{g}_{12}(t)\Delta_{12}(t)]/[g_{12}^2(t)+\Delta_{12}^2(t)]
$
should be applied simultaneously. Given the interaction strength between the two qutrits should be real, such a driving can not be realized directly. However, this difficulty can be solved by applying a $z$-rotation~\cite{STA3}
\begin{eqnarray}
U_z=\left(\begin{array}{cc}
e^{-i\varphi/2}&0\\
0&e^{i\varphi/2}
\end{array}\right), \,\, \varphi(t)=\arctan(\frac{J(t)}{g_{12}(t)}),
\end{eqnarray}
to the system, which delivers the Hamiltonian $H_2(t)+H_2'(t)$ to
\begin{eqnarray}\label{STA2}
H_{2s}&=&U_z(H_2(t)+H_2'(t))U_z^{\dag}-iU_z\dot{U}_z^{\dag}\nonumber\\
&=&\frac{\hbar}{2}[J_s(t)(|e,g\rangle\langle g,e|+|g,e\rangle\langle e,g|)\nonumber\\
&+&2\Delta_s(t)|g,e\rangle\langle g,e|].
\end{eqnarray}
Here, $J_s(t)=\sqrt{J^2(t)+g_{12}^2(t)}$ is real and $\Delta_s(t)=\Delta_{12}(t)-\dot{\varphi}(t)/2$.

Next, we transfer the populations of the qutrits to the two cavities for generating the NOON state. To do this, we switch off the qutrit-qutrit coupling and apply the counter-diabatic driving
\begin{eqnarray}
H_{3i}(t)&=&\frac{\hbar}{2}[\Omega_i^{ef}(t)(|f\rangle_i\langle e|_i+|e\rangle_i\langle f|_i)\nonumber\\
&+&i\tilde{\Omega}_i^{ef}(|f\rangle_i\langle e|_i-|e\rangle_i\langle f|_i)\nonumber\\
&+&2\Delta_i^{ef}(t)|f\rangle_i\langle f|_i],\,\,i=1,2
\end{eqnarray}
to the qutrits simultaneously.
After this, the state of the two qutrits reads $\frac{1}{\sqrt{2}}(|f,g,0,0\rangle+|g,f,0,0\rangle)$. Then, we swap the populations of the qutrits to the respectively coupled resonators. For example, to implement the population of the state $|f_1\rangle$ to the resonator a, we apply the driving
\begin{eqnarray}
H_{41}(t)&=&\frac{\hbar}{2}[g_1'(t)(|e\rangle_1\langle f|_1a^{\dag}+|f\rangle_1\langle e|_1a)\nonumber\\
&+&2\Delta_{1a}^{ef}(t)|f\rangle_1\langle f|_1],
\end{eqnarray}
and set the mixing angle $\theta_4(t)=\arctan[g_1'(t)/\Delta_{1a}^{ef}(t)]/2$ to be changed from zero to $\pi/2$. Note that a counter-diabatic driving
\begin{equation}
H_{41}'(t)=\frac{\hbar}{2}[iG'(t)(|e\rangle_1\langle f|_1a^{\dag}-|f\rangle_1\langle e|_1a)],
\end{equation}
with $
G'(t)=[g_1'(t)\dot{\Delta}_{1a}^{ef}(t)-\dot{g}_1'(t)\Delta_{1a}^{ef}(t)]/[(g_1'(t))^2+(\Delta_{1a}^{ef}(t))^2]
$
is required to apply simultaneously for avoiding the potential non-adiabatic transition. The difficulty of the coupling strength in this auxiliary driving being not real can be similarly solved by applying the following transformation
\begin{eqnarray}
U_{41}=\left(\begin{array}{cc}
e^{-i\varphi_2/2}&0\\
0&e^{i\varphi_2/2}
\end{array}\right),
\end{eqnarray}
which deliver the driving $H_{41}(t)+H'_{41}(t)$ to the experimentally realizable driving
\begin{eqnarray}\label{STA3}
\tilde{H}_{41}(t)&=&\frac{\hbar}{2}[\tilde{g}_1'(t)(|e\rangle_1\langle f|_1a^{\dag}+|f\rangle_1\langle e|_1a)\nonumber\\
&+&2\tilde{\Delta}_{1a}^{ef}(t)|f\rangle_1\langle f|_1],
\end{eqnarray}
with $\varphi_2(t)=\arctan(g_1'(t)/G'(t))$, $\tilde{g}_1'(t)=\sqrt{(g_1'(t))^2+G'(t)}$ and $\tilde{\Delta}_{1a}^{ef}(t)=\Delta_{1a}^{ef}(t)-\dot{\varphi_2}(t)/2$.
Specifically, the AE-type drivings:
$g_1'(t)=G_1\textrm{sech}(\pi t/2T_1)$, and $\Delta_{1a}^{ef}(t)=(2\beta'^2t_0/\pi)\tanh(\pi t/2T_1)$ can be used to deliver the above deterministic population transfers beyond the adiabatic limit.
Similar operations can be performed to the qutrit 2 and the resonator b, and then the state of system is deterministically evolved from $\frac{1}{\sqrt{2}}(|f,g,0,0\rangle+|g,f,0,0\rangle)$ to $\frac{1}{\sqrt{2}}(|e,g,1,0\rangle+|g,e,0,1\rangle)$. Repeating the above operations $N-1$ times,
the following deterministic evolution
\begin{eqnarray}
\frac{1}{\sqrt{2}}(|e,g,1,0\rangle+|g,e,0,1\rangle)\rightarrow\frac{1}{\sqrt{2}}
(|e,g,N,0\rangle+|g,e,0,N\rangle),
\end{eqnarray}
can be implemented.

Finally, The desired NOON state of the two cavities can be generated finally by driving the two qutrits into their ground state $|g,g\rangle$, i.e., the inverse operation of the previous one for preparing the Bell state.

\begin{figure}[t]
\begin{center}
\subfigure[]{\includegraphics[width=0.40\textwidth]{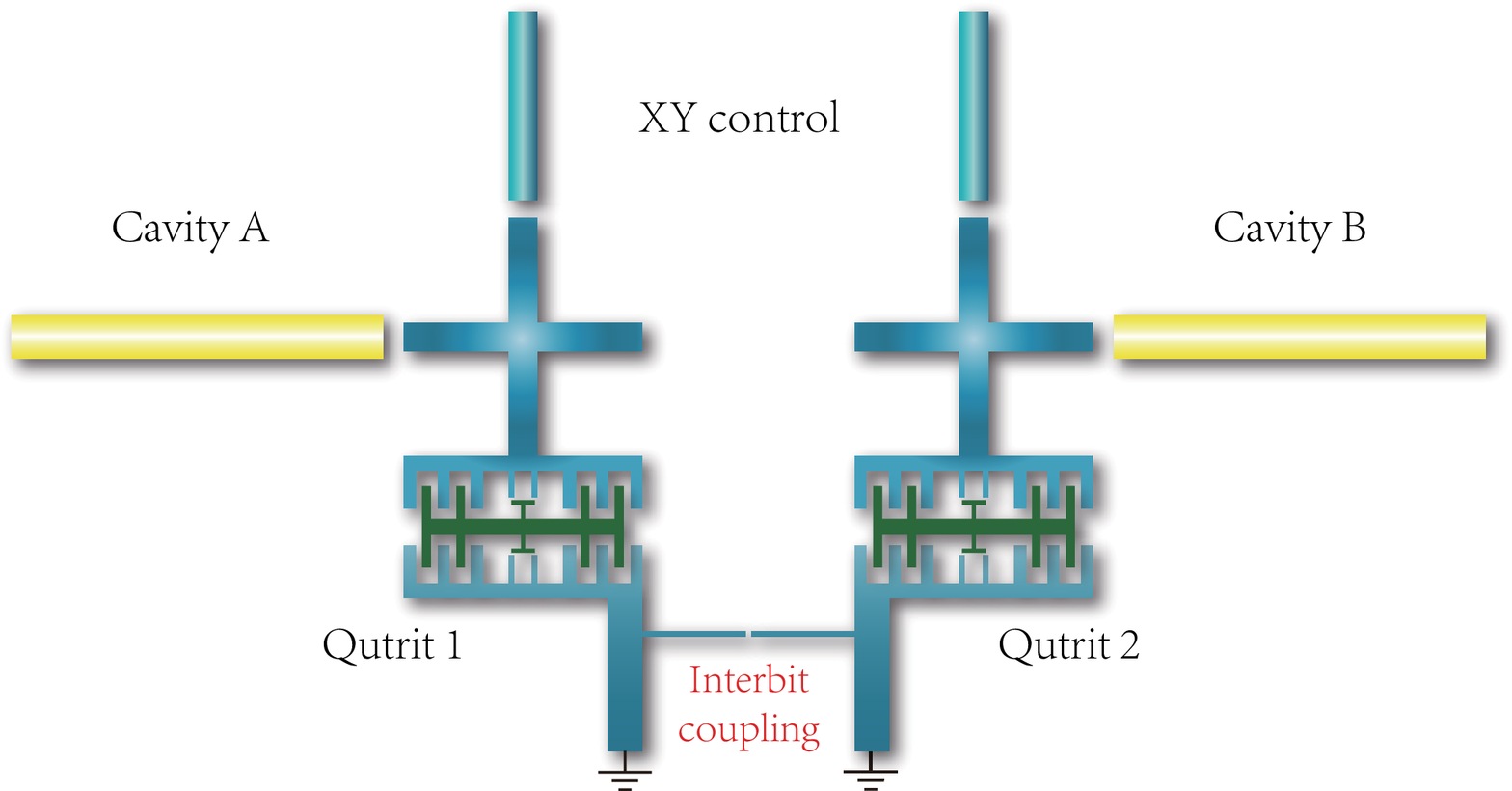}}
\subfigure[]{\includegraphics[width=0.20\textwidth]{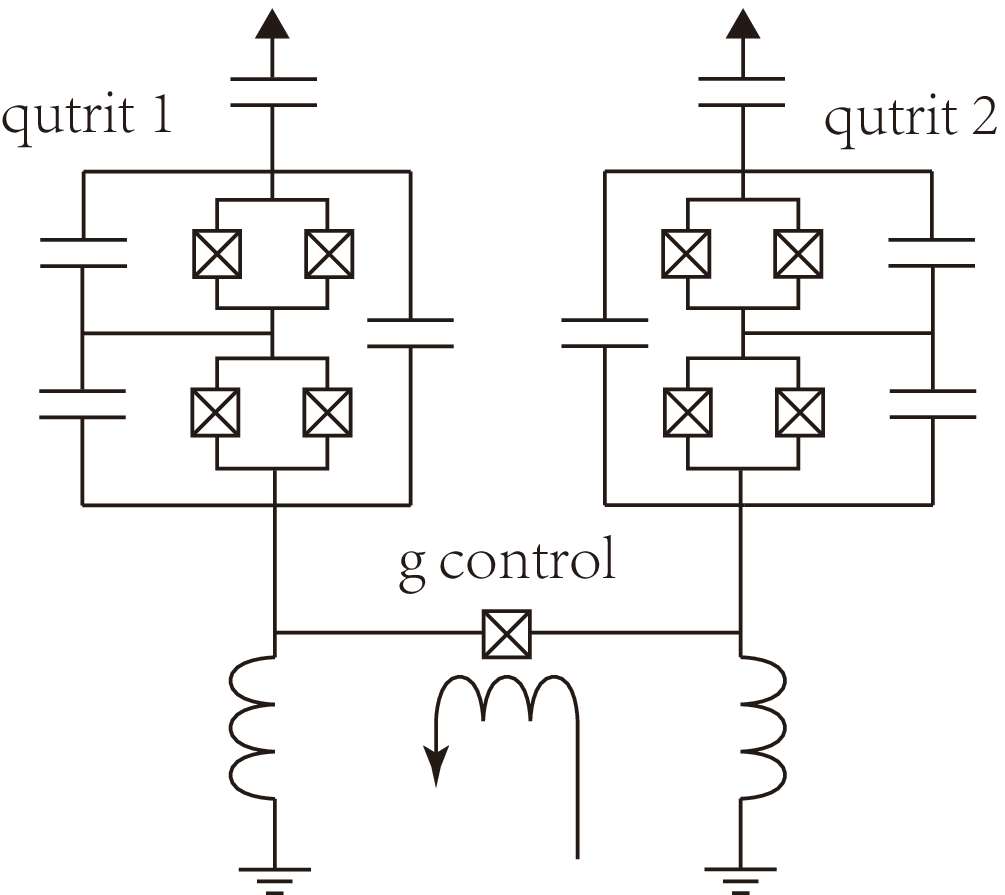}}
\subfigure[]{\includegraphics[width=0.20\textwidth]{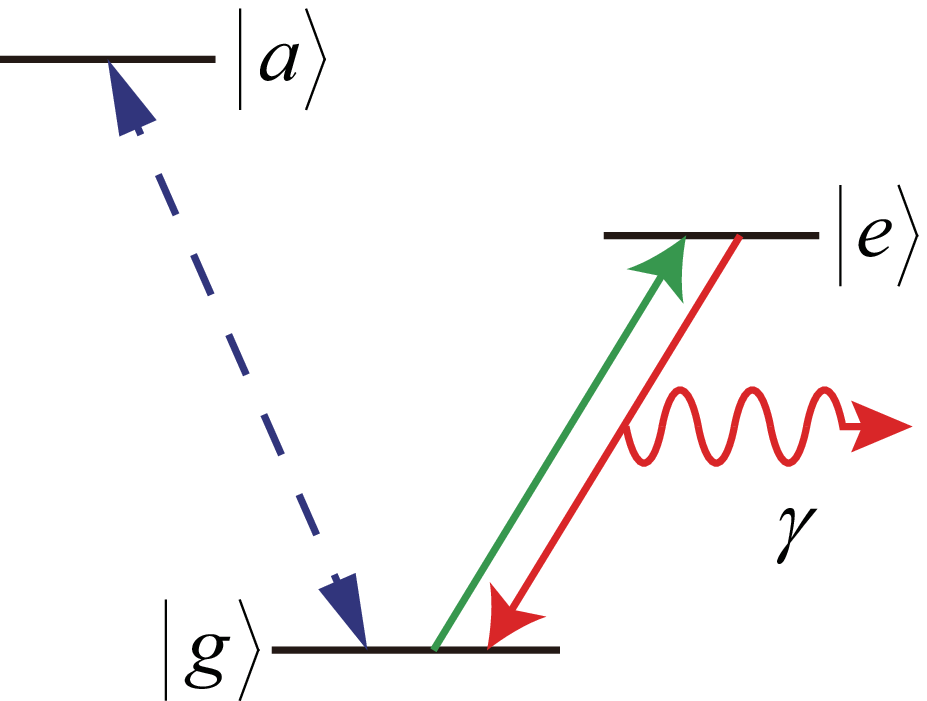}}
\end{center}
\caption{(Color online) (a) Schematic of the superconducting NOON state generator (not to scale) and (b) circuit model of the device. Two qutrits are capacitively coupled with two coplanar waveguide resonators, and inductively coupled with each other by a Josephson junction. The states of qutrits can be independently controlled by XY control and Z control (no shown in figure) circuit. (c) The $V$-type energy level schematic of TCQ, the transition between state $g\leftrightarrow e$ is used to generate photons in the resonator.
}
\end{figure}

{\it Demonstrations with superconducting circuit devices.---}
The previous proposal is quite generic and can be demonstrated with various cavity quantum systems. A promise candidate is the circuit quantum electrodynamical system with tunable qutrit-qutrit and qutrit-cavity interactions. The designed device is schematically shown in Fig.~1, wherein two cavities are coupled together by a gmon~\cite{gmon1,gmon2,gmon3} generated by using tunable-coupling qutrits (TCQs)~\cite{TCQ1,TCQ2,TCQ3} to replace the usual transmons.
Here, each of the TCQs can also be treated as a three-island transmon described by the Hamiltonian $H_T=\sum_{\pm}[4E_{C\pm}(n_{\pm}-n_{g\pm})^2-E_{J\pm}\cos(\varphi_{\pm})]
+4E_In_+n_-$. Here, $n_{+(-)}$ and $n_{g+(-)}$ represent the number of Cooper pairs and offset gate charge in the upper (lower) island, $\varphi_{+(-)}$ are the relevant phase differences; $E_{C\pm}$, $E_{J\pm}$ and $E_I$ represent the charge-, Josephson- and interaction energy, respectively.
In fact, the TCQ can be further modeled as two coupled anharmonic oscillators with annihilation operators $c_{\pm}$ under the condition $E_{J\pm}/E_{C\pm}\gg 1$, and thus  $\bar{H}_T=\sum_{\pm}\hbar[\omega_{\pm}+\delta_{\pm}(c_{\pm}^{\dag}c_{\pm}-1)/2]c_{\pm}^{\dag}c_{\pm}
+\hbar J(c_+c_-^{\dag}+c_+^{\dag}c_-)$ with $\omega_{\pm}=\sqrt{8E_{J\pm}E_{C\pm}}/\hbar-E_{C\pm}/\hbar$, $\delta_{\pm}=-E_{C\pm}/\hbar$ and $J=E_I(E_{J+}E_{J-}/E_{C+}E_{C-})^{1/4}/\sqrt{2}\hbar$.
Obviously, by using the unitary transformation $D(\lambda)=\exp[\lambda(t)(c_+c_-^{\dag}-c_+^{\dag}c_-)]$ this Hamiltonian can be diagonalized as
\begin{eqnarray}
\frac{\tilde{H}_T}{\hbar}=\sum_{\pm}[\frac{\tilde{\delta}_{\pm}}{2}(\tilde{c}_{\pm}^{\dag}\tilde{c}_{\pm}
-1)+\tilde{\omega}_{\pm}]\tilde{c}_{\pm}^{\dag}\tilde{c}_{\pm}+\tilde{\delta}_c\tilde{c}_+^{\dag}\tilde{c}_+\tilde{c}_-^{\dag}\tilde{c}_-, \end{eqnarray}
with $\tilde{c}_{\pm}^{\dag}=c_{\pm}^{\dag}\cos(\lambda)\mp c_{\mp}\sin(\lambda)$; $\lambda(t)=\tan^{-1}(2J/\zeta)/2+\theta$, $\zeta=\omega_+-\omega_--(\delta_+-\delta_-)/2$, and $\theta=\pi$ for $\zeta>0$; $\theta=\pi/2$ for $\zeta<0$.
Physically, the TCQ described above acts as a $V$-type qutrit formed by the levels (see Fig.~1(c)): $|g\rangle=|0\rangle_+|0\rangle_-$, $|e\rangle=\tilde{c}_+^{\dag}|g\rangle$, and $|a\rangle=\tilde{c}_-^{\dag}|g\rangle$.
Note that the formal qutrit-cavity coupling strengths  $\tilde{g}_{\pm}=g_{\pm}\cos(\lambda)\mp g_{\mp}\sin(\lambda)$ can be  frequency-independent, and only the parameter $\tilde{g}_+$ can be tuned to zero due to the negative $J$~\cite{TCQ1}. Thus, the transition between $|g\rangle$ and $|e\rangle$ can be utilized to excite the coupled resonator. Given the coupling between the distant qutrits mediated by an auxiliary cavity demonstrated in the previous configuration~\cite{N8} is typically weak, implying the relative-long operational durations, we alternatively introduce a tunable Josephson junction to generate an sufficiently-strong inter-bit coupling $S(\theta)\varphi_{1-}\varphi_{2-}$. The present coupling strength $S$ can also be tuned by modulating $\theta$, the phase difference across the coupling junction. Specifically,
$
S(\theta)\varphi_{1-}\varphi_{2-}= g_{12}(\theta)(c_-^{\dag}+c_-)(d_-^{\dag}+d_-)
=\tilde{g}_{12}(\theta)(\tilde{c}_-^{\dag}+\tilde{c}_-)
(\tilde{d}_-^{\dag}+\tilde{d}_-)
\approx\tilde{g}_{12}(\theta)(\tilde{c}_-^{\dag}\tilde{d}_-+\tilde{c}_
-\tilde{d}_-^{\dag}),
$
with $c$ and $d$ representing the annihilation operators of the qutrit 1 and 2, respectively. The last step is due to the usual rotating wave approximation. Consequently, the tunable qutrit-qutrit coupling $\hbar g_{12}(t)(|g,a\rangle\langle a,g|+|a,g\rangle\langle g,a|)$ desired similarly in Eq.~(8) can be obtained.

\begin{figure*}[t]
\begin{center}
\subfigure[]{\includegraphics[width=0.25\textwidth]{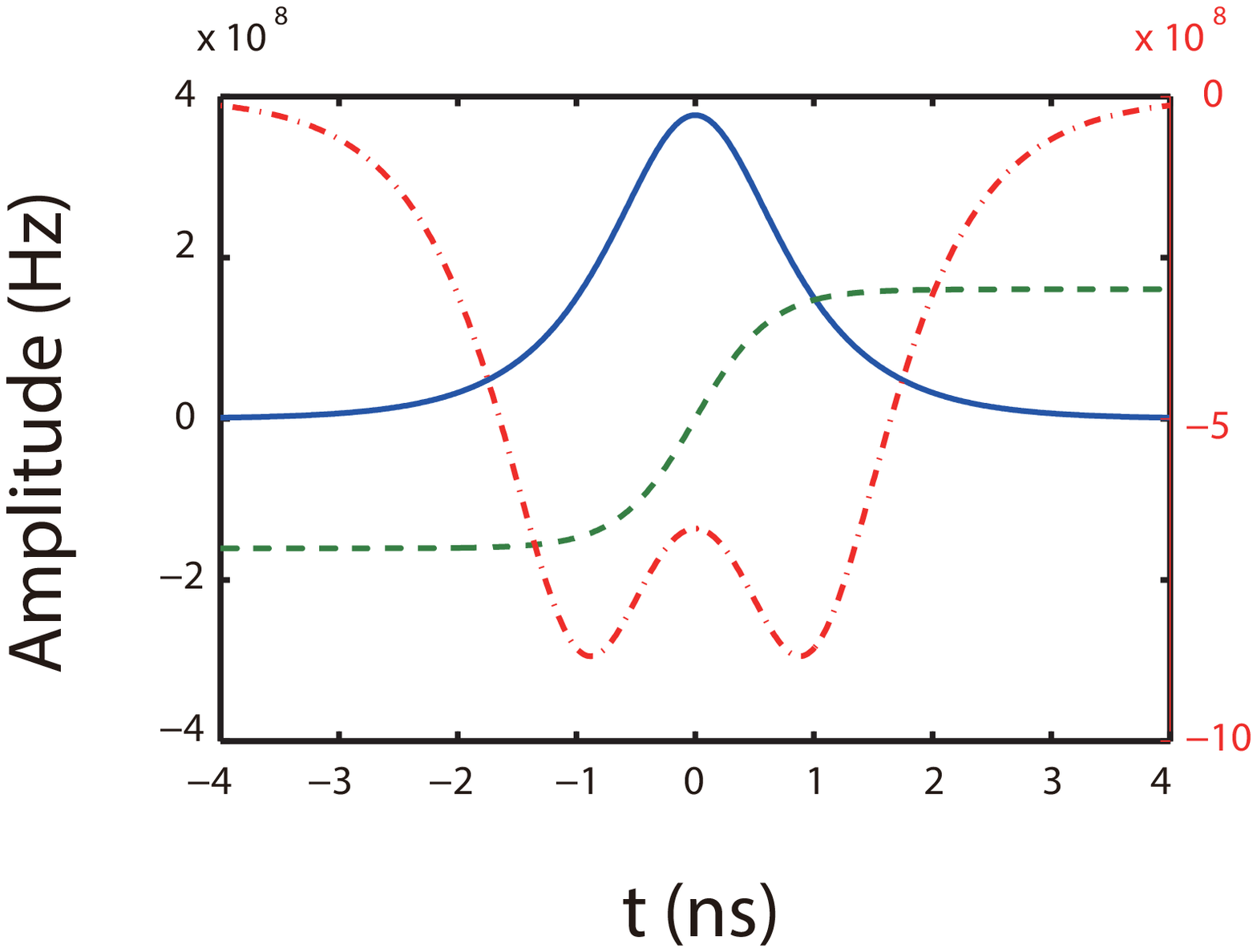}}
\subfigure[]{\includegraphics[width=0.25\textwidth]{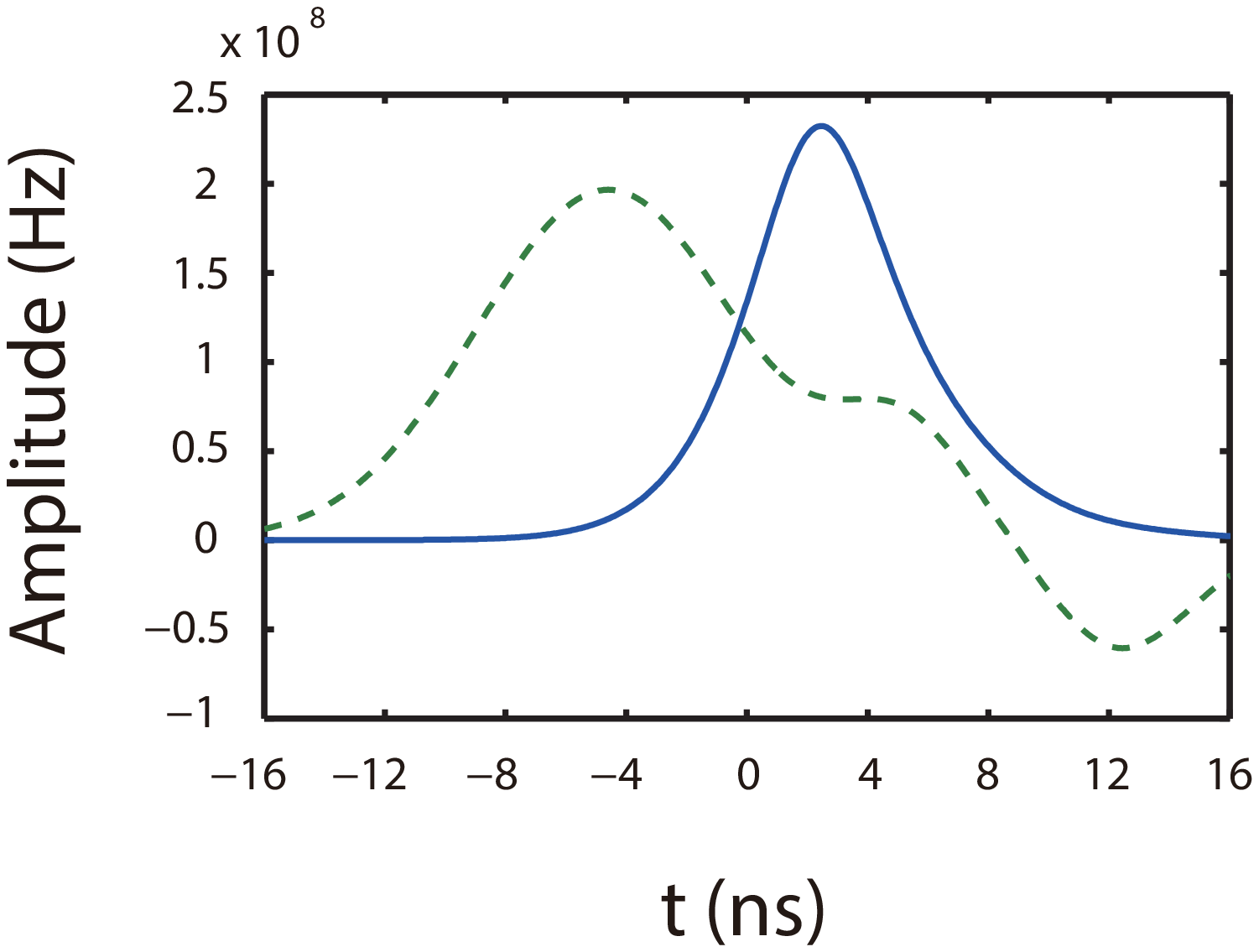}}
\subfigure[]{\includegraphics[width=0.25\textwidth]{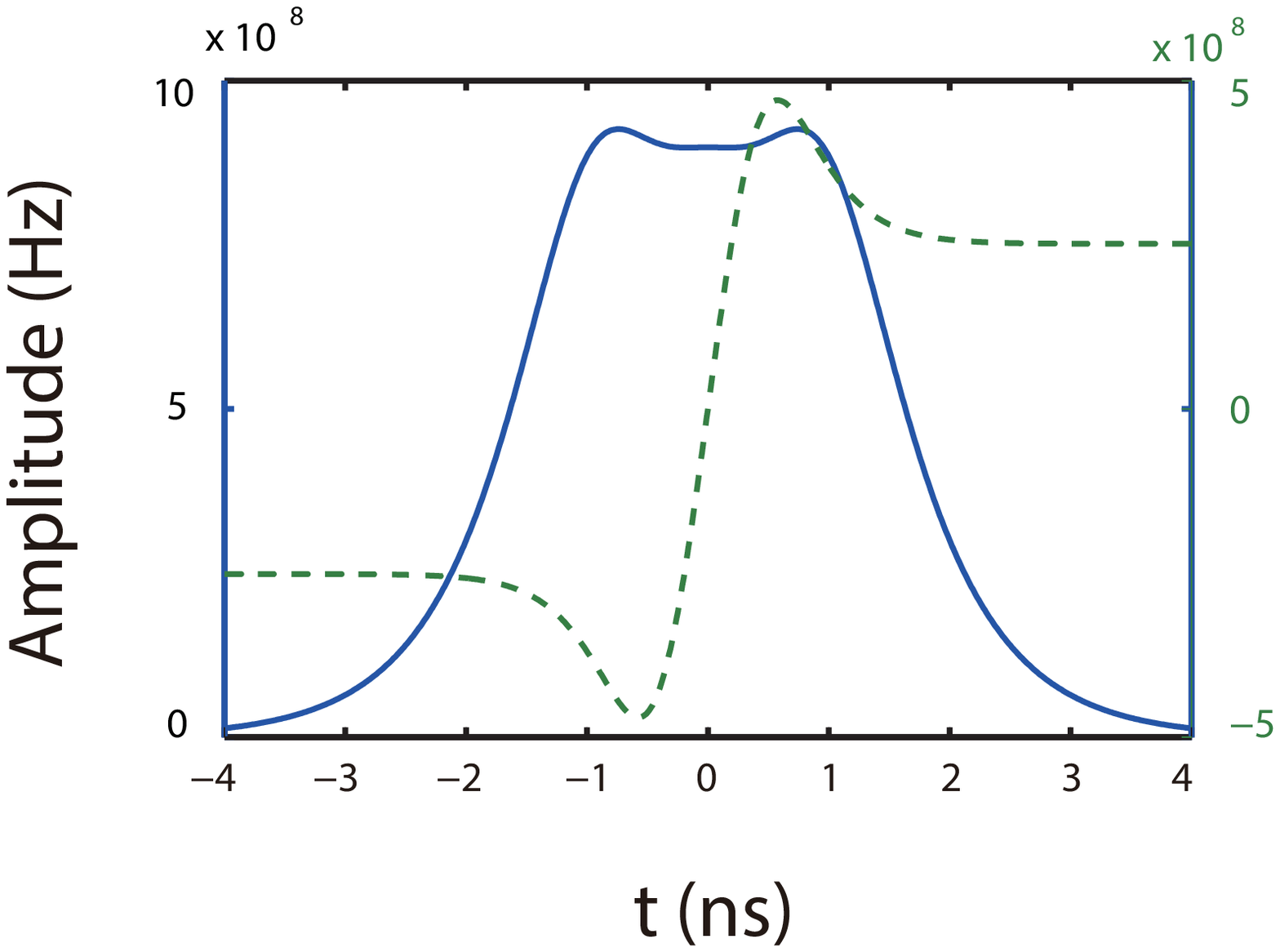}}
\subfigure[]{\includegraphics[width=0.25\textwidth]{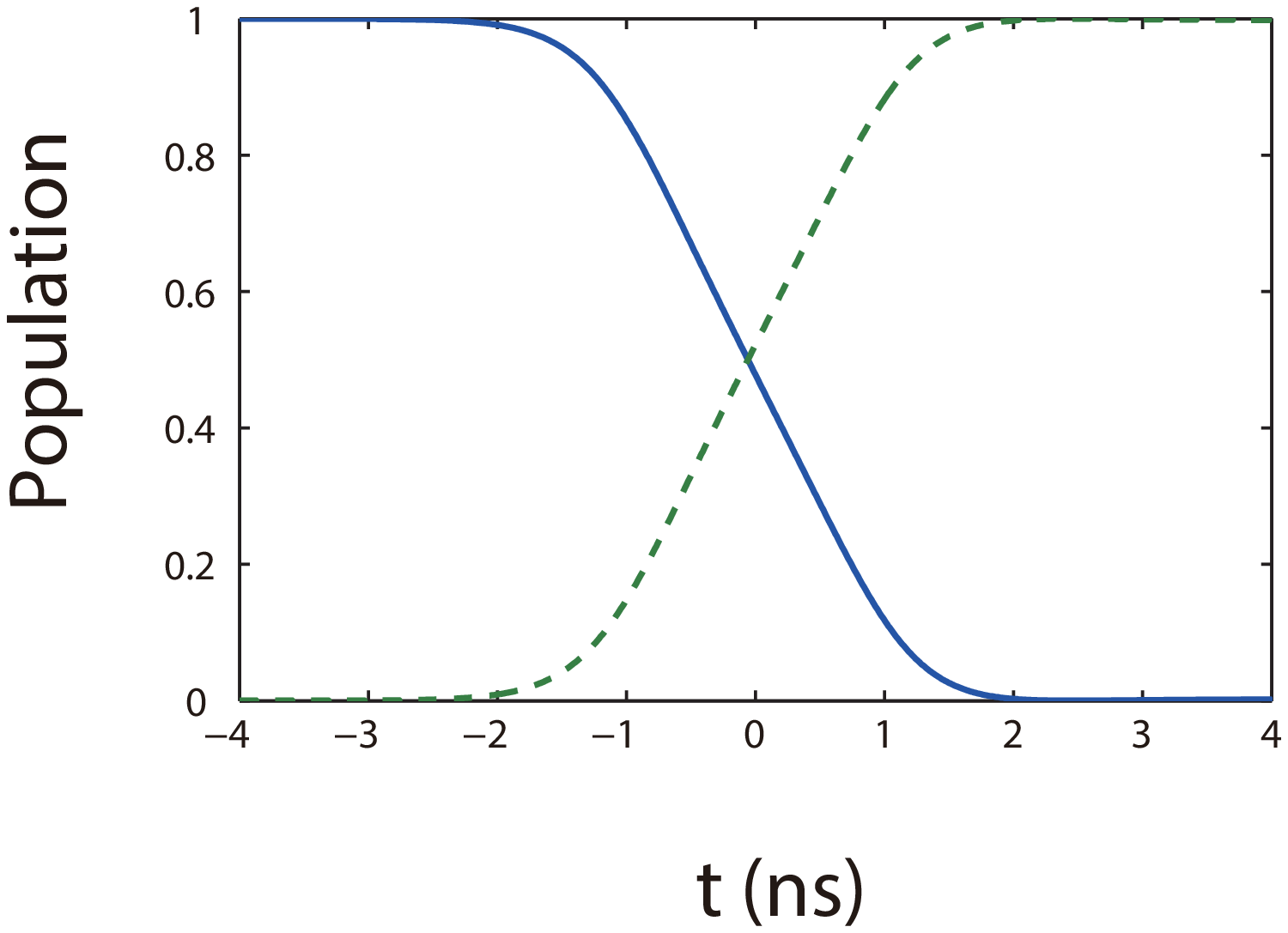}}
\subfigure[]{\includegraphics[width=0.25\textwidth]{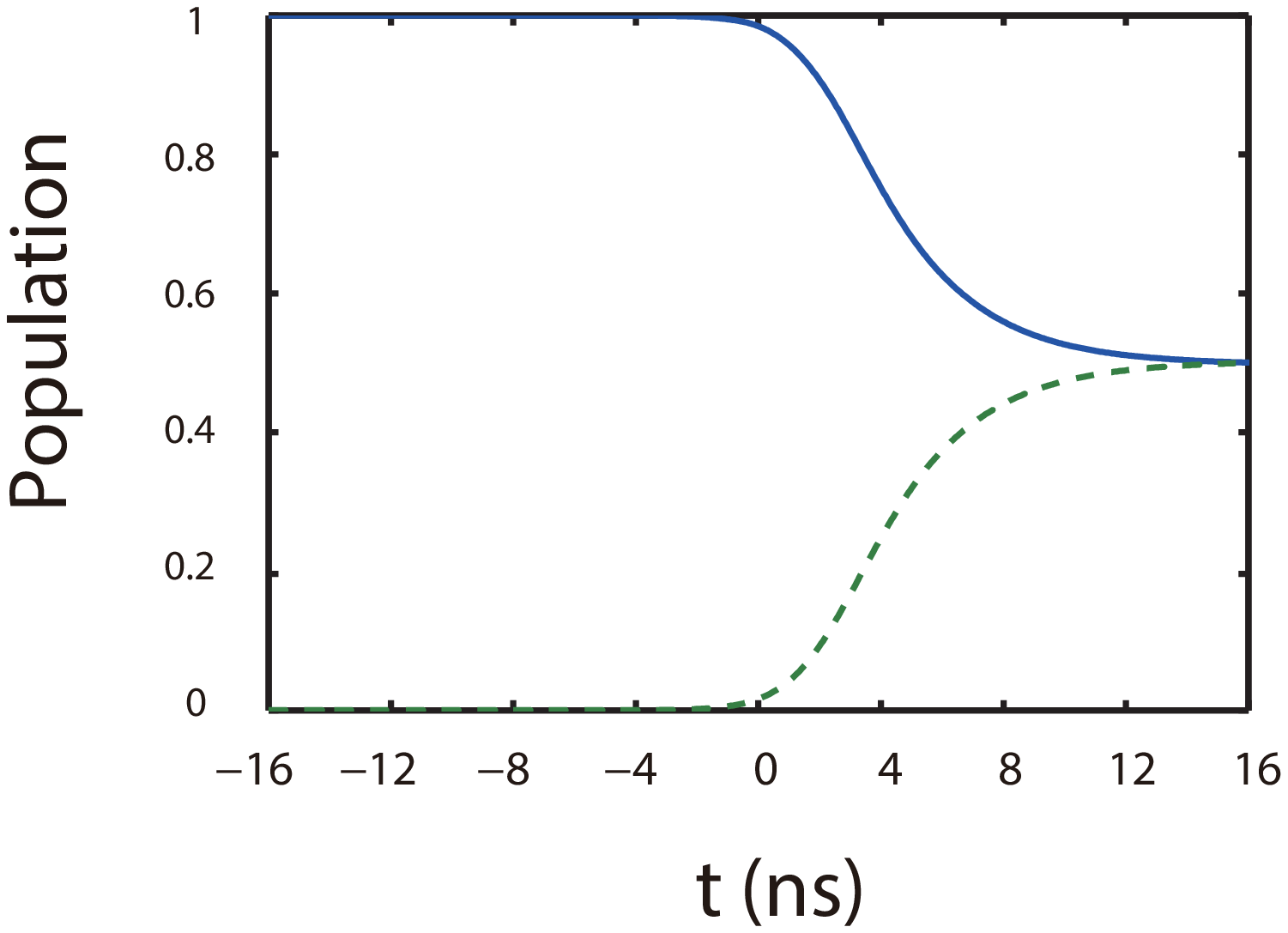}}
\subfigure[]{\includegraphics[width=0.25\textwidth]{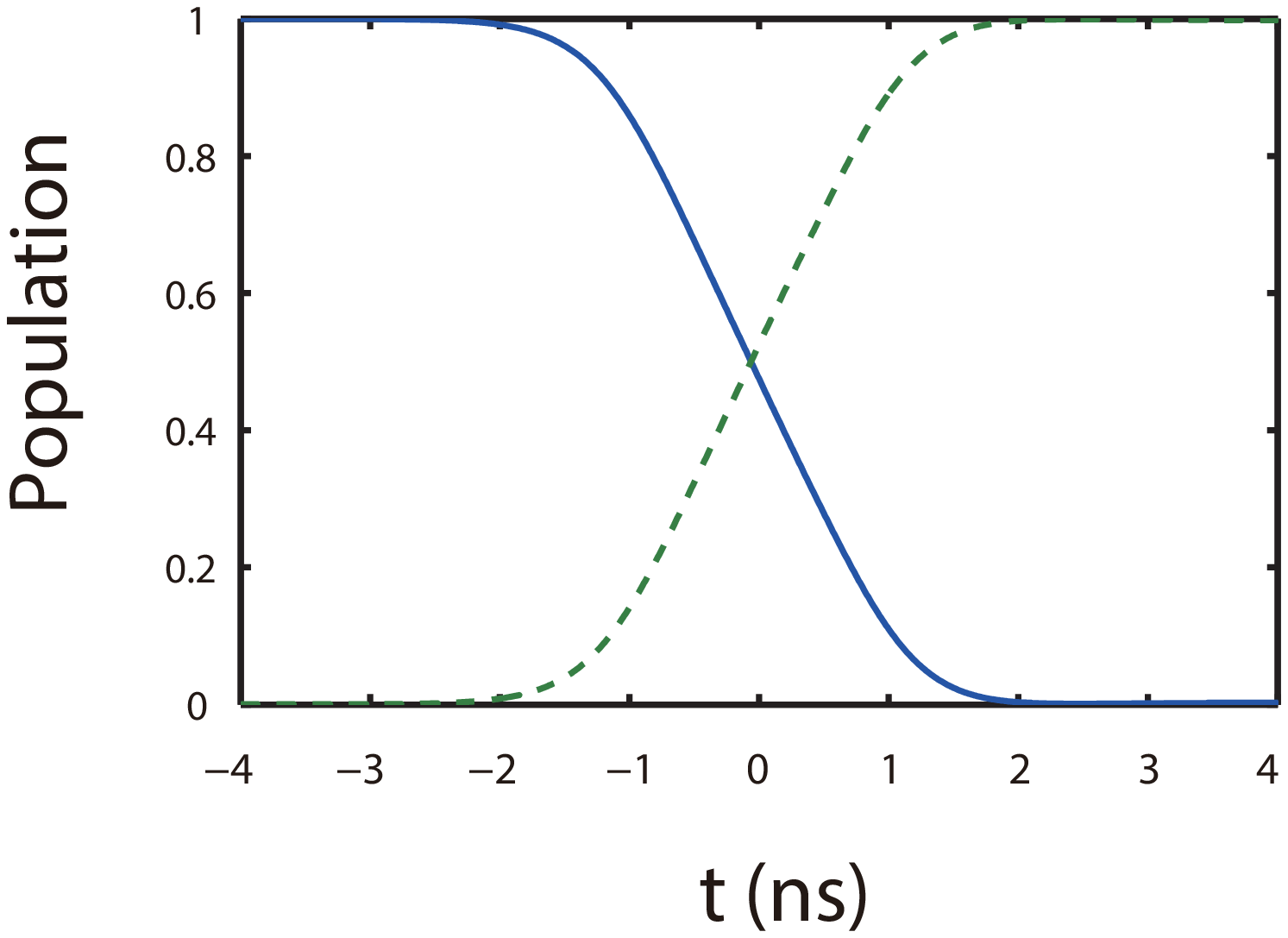}}
\end{center}
\caption{(Color online) Implementation of transition from: (a) and (d), $|g,g,0,0\rangle$ to $|a,g,0,0\rangle$; (b) and (e), $|a,g,0,0\rangle$ to $(|g,a,0,0\rangle+|a,g,0,0\rangle)/\sqrt{2}$; (c) and (f), $|e,a,k,0\rangle$ to $|a,e,k+1,0\rangle$.
Here, (a), (b) and (c) are the operated drivings, with (a) the AE drivings $\Omega_1(t)$ (solid blue line, left axis) and $\Delta_1(t)$ (dash green line, left axis), and the counter-diabatic driving $\Omega_1'(t)$ (dot-dash red line, right axis); (b) coupling strength $J_s(t)$ (solid blue line) and frequency detuning $\Delta_s(t)$ (dash green line); (c) STA drivings: $g_1'(t)$ (solid blue line) and $\Delta_{1a}^{ge}(t)$ (dash green line).
(d), (e) and (f) are the relevant population transfer in the STA process, where the solid blue lines denote the population of the state $|g,g,0,0\rangle$, $|a,g,0,0\rangle$ and $|e,a,k,0\rangle$, and the dash green lines denote the population of the state $|a,g,0,0\rangle$, $|g,a,0,0\rangle$ and $|g,a,k+1,0\rangle$, respectively.}
\end{figure*}

\begin{figure*}[t]
\begin{center}
\subfigure[]{\includegraphics[width=0.25\textwidth]{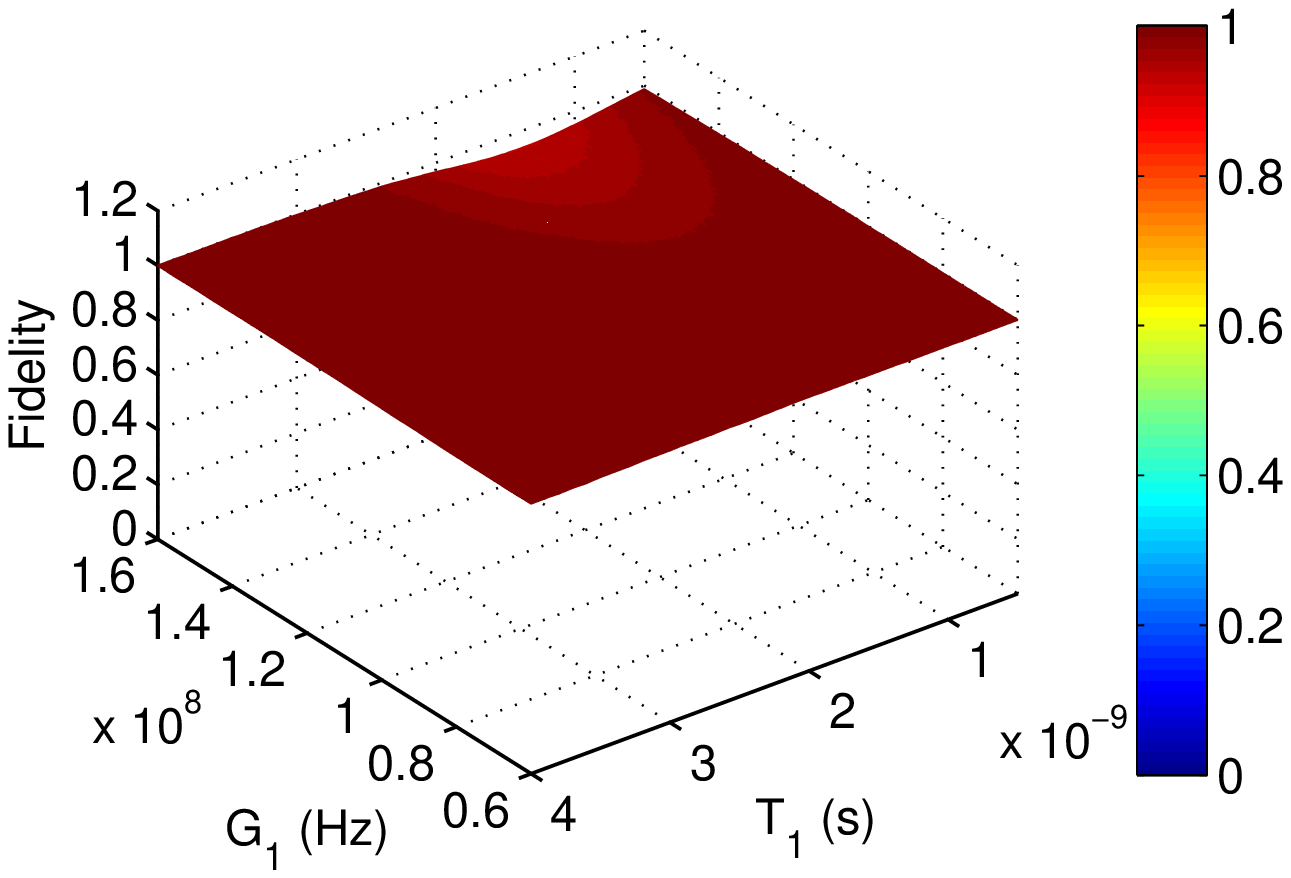}}
\subfigure[]{\includegraphics[width=0.25\textwidth]{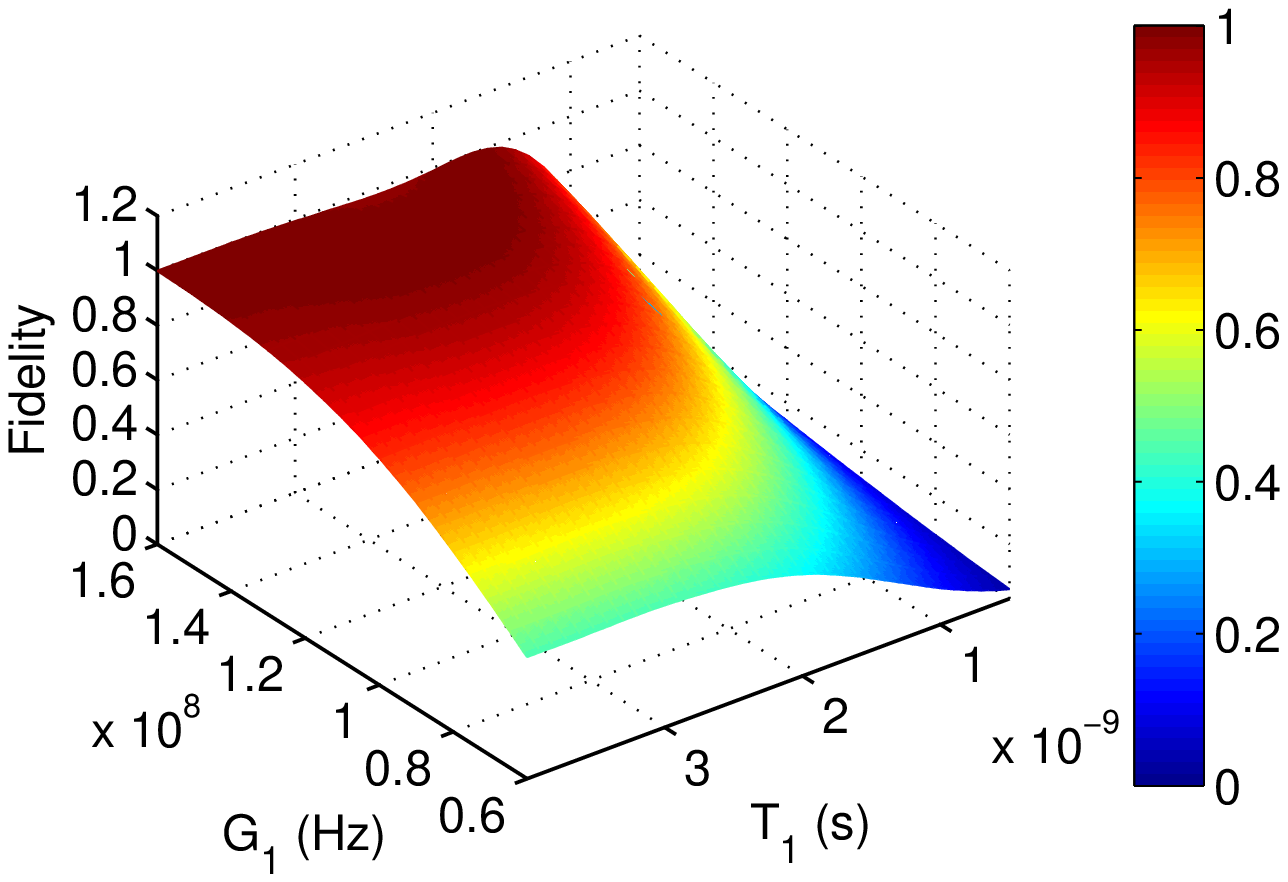}}
\subfigure[]{\includegraphics[width=0.25\textwidth]{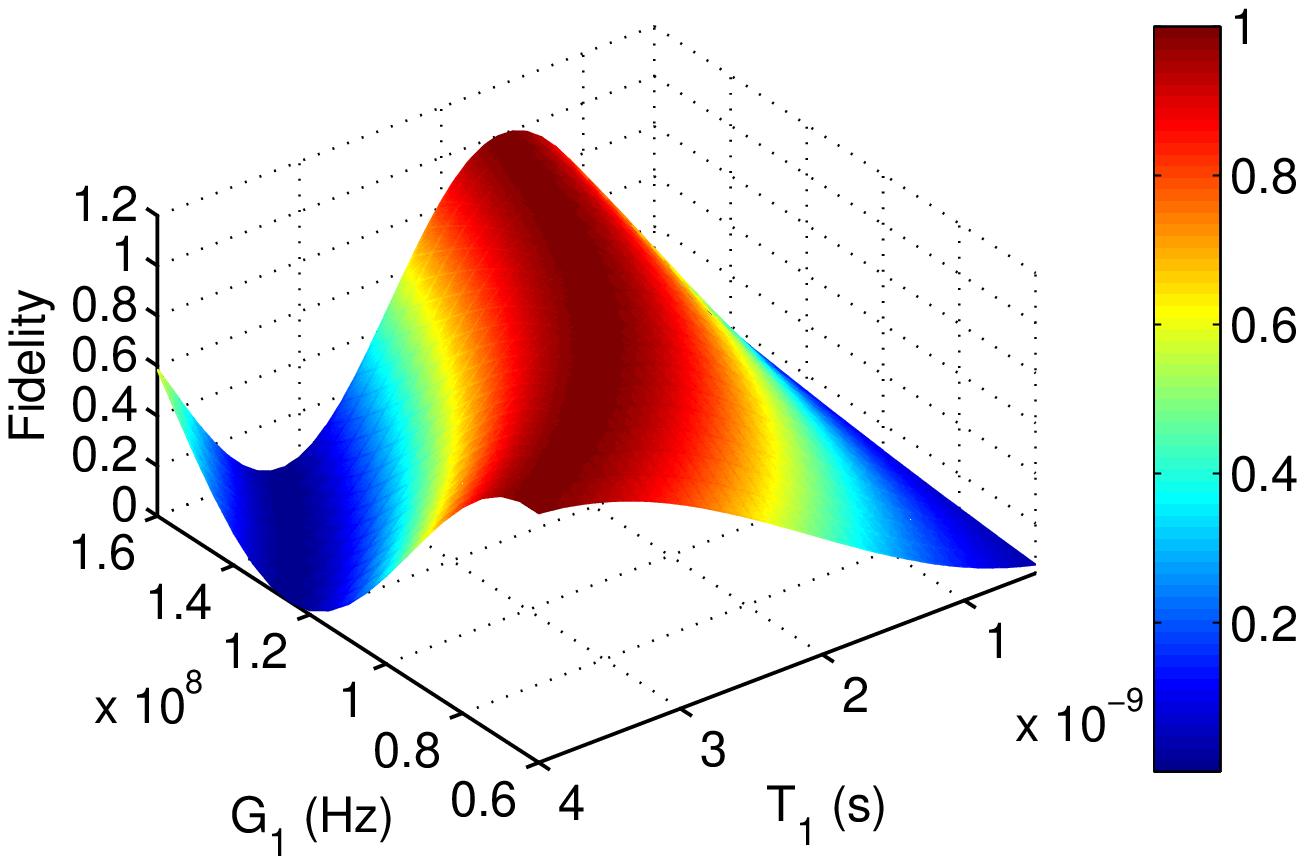}}
\end{center}
\caption{(color online) Fidelity of the implementation of transition from $|e,a,k,0\rangle$ to $|g,a,k+1,0\rangle$ by: (a) the STA technique, (b) the adiabatic passage technique, and (c) the Rabi-ocsillation technique. }
\end{figure*}

We now show how to demonstrate the deterministic generation of the NOON state with the above superconducting circuit. As the present qutrit is $V$-type, rather than the ladder-type qutrits used in our generic proposal described in the last section, the procedure for the generation of the desired NOON state should be slightly changed as:
a) $|g,g,0,0\rangle\rightarrow|a,g,0,0\rangle\rightarrow\frac{1}{\sqrt{2}}(|g,a\rangle+|a,g\rangle)
\otimes|0,0\rangle$; b) $\frac{1}{\sqrt{2}}(|g,a,k,0\rangle+|a,g,0,k\rangle)\rightarrow
\frac{1}{\sqrt{2}}(|e,a,k,0\rangle+|a,e,0,k\rangle)\rightarrow\frac{1}{\sqrt{2}}
(|g,a,k+1,0\rangle+|a,g,0,k+1\rangle)$, with $k=0,1,2,...N-1$; and c) $\frac{1}{\sqrt{2}}(|g,a,N,0\rangle+|a,g,0,N\rangle)\rightarrow
\frac{1}{\sqrt{2}}|g,g\rangle\otimes(|N,0\rangle+|0,N\rangle)$.
The effectiveness of such a procedure is verified by our numerical simulations. For example, in Fig.~2(a) and (d) we show the population transfer: $|g,g,0,0\rangle\rightarrow|a,g,0,0\rangle$, could be deterministically realized by using the Hamiltonian Eq.~(\ref{STA1}) with  the relevant counter-diabatic driving Eq.~(\ref{STA1a}). In the figures the relevant parameters are chosen as $t_0=1$ns, $\Omega_0=2\pi\times60$MHz, and $\beta=2\pi\times80$MHz. One can see that, the desired population passage is performed almost perfectly with significantly-high fidelity and the passage duration is less than 6 nanoseconds.
Also, Fig.~2(b) and (e) show how the Bell state $(|g,a,0,0\rangle+|a,g,0,0\rangle)/\sqrt{2}$ could be prepared deterministically by using the Hamiltonian (\ref{STA2}), with parameters $T_0=5$ns, $G_0=2\pi\times10$MHz, $\Delta_0=2\pi\times30$MHz, $m=1.25$, and $\tau=4$ns. Note that the present duration is nearly $16$ns restricted mainly by the maximum of the inter-qutrit coupling strength which was measured as $2\pi\times55$MHz for the demonstrated gmon~\cite{gmon1}.
Finally, Fig.~2(c) and (f) show how to swap the population from the qutrit to the resonator by using the Hamiltonian (\ref{STA3}) in $5$ns, with parameters $T_1=1$ns, $G_1=2\pi\times90$MHz, and $\beta'=2\pi\times100$MHZ.

We now investigate the fidelity of the above generations with the parameter variations. Without loss of the generality, we focus the last part in the step b) as this step has been repeat several times for the desired NOON state  generation. Again, we use the AE drivings: $g_1'(t)=G_1\textrm{sech}(\pi t/2T_1)$, and $\Delta_{1a}^{ge}(t)=(2\beta'^2t_0/\pi)\tanh(\pi t/2T_1)$, and the parameters $G_1$ is changed from $60$MHz to $160$MHz, and $T_1$ is changed from $0.5$ns to $4$ns. With these common drivings we compare the fidelities of the population passages by the STA-, the APP- and the RO techniques in Fig.~3, respectively. It is shown that the STA technique can achieve high-fidelity in any case but the APP one only works well for the sufficiently-long process time. Of course, the RO technique works only with a $\pi/2$ pulse integral area. This indicates that the STA technique delivers more robustness with significantly-high fidelity for a common duration.

{\it Conclusion and Discussions.---}
In conclusion, we theoretically demonstrate that the photonic NOON states between two cavities can be deterministically generated by using the STA technique. Due to the duration of such a generation is significantly shorter than that by the APP, and also more robustness than that by the usual RO to the fluctuations of the pulse parameters, the fidelity of the NOON state generation should be sufficiently high and feasible.

The generic proposal has been demonstrated specifically with the experimental superconducting circuits. In order to speedup the generation for a definite coherent time we used the gmon configuration to realize the direct controllable qutrit-qutrit interaction and the tunable qutrit-cavity couplings. Typically, all the coupling strengths in our device configuration are tunable independently, and thus some dispensable couplings can be effectively switched off. For example, once the Bell state has been prepared, the inter-bit coupling can be tuned to zero so that the system divides to two qutrit-resonator subsystem. More importantly, only two excited states of the $V$-type atom are used in our experimental demonstrations, and thus the stronger decoherence of higher levels should not influence the fidelity of the proposed NOON state generation.
Hopefully, the scheme for generating the desired NOON state with high fidelity can be directly used in various quantum metrology applications.

{\it Acknowledgements.---}
This work was supported in part by the NSFC grant Nos. 11174373, 61301031, U1330201.


\begin{thebibliography}{0}

\bibitem{N1} B. L. Higgins1, D. W. Berry, S. D. Bartlett, H. M. Wiseman and G. J. Pryde, Nature 450, 393 (2007).

\bibitem{N2} V. Giovannetti, S. Lloyd and L. Maccone, Nature Photonics 5, 222 (2011).

\bibitem{N3} T. Nagata, R. Okamoto, J. L. O'Brien, K. Sasaki, and S. Takeuchi, Science 316, 726 (2007).

\bibitem{N4} A. N. Boto, P. Kok, D. S. Abrams, S. L. Braunstein, C. P. Williams, and J. P. Dowling, Phys. Rev. Lett. 85, 2733 (2000).

\bibitem{N5} I. Afek, O. Ambar, and Y. Silberberg, Science 328, 879 (2010).

\bibitem{N6} Y. Israel, I. Afek, S. Rosen, O. Ambar, and Y. Silberberg, Phys. Rev. A 85, 022115 (2012).

\bibitem{N7} S. T. Merkel, and F. K Wilhelm, New J. Phys. 12, 093036 (2010).

\bibitem{N8} H. Wang, M. Mariantoni, R. C. Bialczak, M. Lenander, E. Lucero, M. Neeley, A. D. O¡¯Connell, D. Sank, M. Weides, J. Wenner, T. Yamamoto, Y. Yin, J. Zhao, J. M. Martinis, and A. N. Cleland, Phys. Rev. Lett. 106, 060401 (2011).

\bibitem{N9} F. W. Strauch, K. Jacobs, and R. W. Simmonds, Phys. Rev. Lett. 105, 050501 (2010).

\bibitem{N10} F. W. Strauch, Phys. Rev. Lett. 109, 210501 (2012).

\bibitem{N11} Q. Su, C. Yang and S. Zheng, Sci. Rep. 4, 3898 (2014).

\bibitem{N12} S. Xiong, Z. Sun, J. Liu, T. Liu, and C. Yang, Opt. Lett. 40, 2221-2224 (2015).

\bibitem{N13} S. S. Ivanov, N. V. Vitanov, and N. V Korolkova, New J. Phys. 15, 023039 (2013).

\bibitem{N14} C. H. Bennett, and David P. DiVincenzo, Nature 404, 247 (2000).%QIP

\bibitem{N15} N. Gisin, and R. Thew, Nature Photonics 1, 165 (2007).%communication

\bibitem{CQED1} A. Blais, R. S. Huang, A. Wallraff, S. M. Girvin, and R. J. Schoelkopf, Phys. Rev. A 69 062320 (2004).

\bibitem{CQED2} A. Blais, J. Gambetta, A. Wallraff, D. I. Schuster, S. M. Girvin, M. H. Devoret, and R. J. Schoelkopf, Phys. Rev. A 75, 032329 (2007).

\bibitem{AP1} K. Bergmann, H. Theuer, and B. W. Shore, Rev. Mod. Phys. 70, 1003 (1998).

\bibitem{AP2} P. Kr\'{a}l, I. Thanopulos, and M. Shapiro, Rev. Mod. Phys. 79, 53 (2007).

\bibitem{AP3} N. V. Vitanov, T. Halfmann, B. W. Shore, and K. Bergmann, Annu. Rev. Phys. Chem. 52, 763 (2001).

\bibitem{AP4} L. F. Wei, J. R. Johansson, L. X. Cen, S. Ashhab, and F. Nori, Phys. Rev. Lett. 100, 113601 (2008).

\bibitem{STA1} M. V. Berry, J. Phys. A 42, 365303 (2009).

\bibitem{STA2} X. Chen, I. Lizuain, A. Ruschhaupt, D. Gu\'{e}ry-Odelin, and J. G. Muga, Phys. Rev. Lett. 105, 123003 (2010).

\bibitem{STA3} M. G. Bason, M. Viteau, N. Malossi, P. Huillery, E. Arimondo, D. Ciampini, R. Fazio, V. Giovannetti, R. Mannella, and O. Morsch, Nat. Phys. 8, 147 (2012).

\bibitem{STA4} M. Demirplak and S. A. Rice, J. Phys. Chem. A 107, 9937 (2003).

\bibitem{STA5} X. Chen, A. Ruschhaupt, S. Schmidt, A. del Campo, D. Gu\'{e}ry-Odelin, and J. G. Muga, Phys. Rev. Lett. 104, 063002 (2010).

\bibitem{STA6} X. xuan, and L. F. Wei, Laser Phys. Lett. 12, 015204 (2015).

\bibitem{STA7} J. Chen, and L. F. Wei, Phys. Rev. A 91,023405 (2015).

\bibitem{AE} L. Allen and J. H. Eberly, Optical Resonance and Two-Level Atoms (Dover, New York, 1987).

\bibitem{gmon1} Y. Chen, C. Neill, P. Roushan, N. Leung, M. Fang, R. Barends, J. Kelly, B. Campbell, Z. Chen, B. Chiaro, A. Dunsworth, E. Jeffrey, A. Megrant, J. Y. Mutus, P. J. J. O¡¯Malley, C. M. Quintana, D. Sank, A. Vainsencher, J. Wenner, T. C. White, M. R. Geller, A. N. Cleland, and J. M. Martinis, Phys. Rev. Lett. 113, 220502 (2014).

\bibitem{gmon2} R. Barends, J. Kelly, A. Megrant, D. Sank, E. Jeffrey, Y. Chen, Y. Yin, B. Chiaro, J. Mutus, C. Neill, P. O¡¯Malley, P. Roushan, J. Wenner, T. C. White, A. N. Cleland, and J. M. Martinis, Phys. Rev. Lett. 111, 080502 (2013).

\bibitem{gmon3} M. R. Geller, E. Donate, Y. Chen, M. T. Fang, N. Leung, C. Neill, P. Roushan, and J. M. Martinis, Phys. Rev. A 92, 012320 (2015).

\bibitem{TCQ1} J. M. Gambetta, A. A. Houck, and A. Blais, Phys. Rev. Lett. 106, 030502 (2011).

\bibitem{TCQ2} S. J. Srinivasan, A. J. Hoffman, J. M. Gambetta, and A. A. Houck, Phys. Rev. Lett. 106, 083601 (2011).

\bibitem{TCQ3} A. Mezzacapo, L. Lamata, S. Filipp, and E. Solano, Phys. Rev. Lett. 113, 050501 (2014).

\end{thebibliography}
\end{document}